\documentclass[prl,reprint,twocolumn,showpacs,preprintnumbers,amsmath,amssymb,
nofootinbib,tightenlines,groupedaddress,superscriptaddress]{revtex4-1} 
\usepackage{graphicx}

\usepackage{mathrsfs}
\usepackage{hyperref}
\usepackage{mathtools}

\usepackage{color}
 \usepackage[utf8]{inputenc}
 \usepackage{booktabs}  

\newcommand{\bea}{\begin{eqnarray}}
\newcommand{\eea}{\end{eqnarray}}
\newcommand{\bean}{\begin{eqnarray*}}
\newcommand{\eean}{\end{eqnarray*}}

\def\Label#1{\label{#1}%
  \smash{\hbox to0pt{\raise1ex\hbox{\tiny[#1]}\hss}}}
  
\renewcommand{\eqref}[1]{eq.~(\ref{#1})}

\begin{document}

\title{Analytic Form of the Three-loop Four-gluon Scattering Amplitudes in Yang-Mills Theory} 

\author{Qingjun Jin}
\email{qjin@itp.ac.cn}
\affiliation{Graduate School of China Academy of Engineering Physics, No. 10 Xibeiwang East Road, Haidian District, Beijing, 100193, China}
\affiliation{CAS Key Laboratory of Theoretical Physics, Institute of Theoretical Physics, \\ Chinese Academy of Sciences, Beijing 100190, China}

\author{Hui Luo}
\email{hluo@gscaep.ac.cn}
\affiliation{Graduate School of China Academy of Engineering Physics, No. 10 Xibeiwang East Road, Haidian District, Beijing, 100193, China}

\date{\today}
\begin{abstract}
We present analytic forms of three-loop four-gluon planar amplitudes in pure Yang-Mills theory in this letter. 
Gauge invariant bases and a set of proper master integrals are chosen such that the amplitudes are explicitly invariant under cyclic permutations of external particles. 
The $D$-dimensional unitarity method and integration-by-parts reductions with cut conditions are performed to determine the coefficients of the master integrals. 
Helicity amplitudes in the conventional dimension regularization scheme are obtained by setting gauge invariant bases to helicity configurations. 
After renormalizing the ultraviolet divergences, remaining divergences of the amplitudes agree with known infrared divergence structures exactly.
Our results provide an essential piece of three-loop amplitudes required for the ${\rm N^3LO}$ order corrections to the production of two jets at hadron colliders.
\end{abstract}

\pacs{12.38.Bx }

\keywords{}

\maketitle

\section{Introduction}

The standard model has been now well-established after Higgs boson confirmed by the Large Hadron Collider (LHC) at CERN.
Along with updates of LHC, its energy and luminosity will both increase, which would lead to higher precision of experimental results.
To search for new physics beyond Standard Model (SM) as well as to explore physics properties of existed particles, theoretical results of scattering cross section to a certain precision are required to be compared with experimental results. 

Take the di-jet production at the LHC as an example, a bump-like feature in the di-jet mass distribution could be an evidence of new physics signal \cite{Aad:2014aqa, Khachatryan:2015sja, ATLAS:2015nsi, Sirunyan:2016iap, Aaboud:2017yvp}, where processes from Quantum Chromodynamics (QCD) serve as the backgrounds.
Furthermore, di-jet system could be directly used to test validity of perturbations of QCD and properties of strong couplings at different energy scales \cite{Chatrchyan:2013txa, ATLAS:2013lla}.
The amplitudes of di-jet production from QCD up to next-to-next-to-leading order (NNLO) have been studied years ago \cite{Popov:2001mn,Glover:2001rd,Bern:2002tk,Anastasiou:2001sv,Anastasiou:2000mv, Glover:2003cm,Bern:2003ck}.
The comparison between theoretical predictions at NNLO and experimental measurements has been carried out lately \cite{Currie:2017eqf, Bellm:2019yyh}.


To prepare for future high luminosity experiments in LHC, at least ten times of current luminosity, further reducing uncertainty in the QCD background of di-jet bump hunting is a necessary task to be accomplished, namely, to compute the $\textrm{N}^3$LO contributions to di-jet theoretic prediction.
Not only is this computation a preliminary for updated LHC experimental measurements, but also a useful element for testing the convergence of QCD perturbations and the running properties of strong couplings.
In this letter, we present analytic forms of the leading color four-gluon three-loop helicity amplitudes in pure Yang-Mills theory, as a first result for di-jet prediction at $\textrm{N}^3$LO.

We compute using d-dimensional unitarity method \cite{Bern:1994zx, Bern:1995db}, gauge invariant bases method \cite{Boels:2017gyc, Boels:2018nrr}  and integration-by-parts (IBP) reduction \cite{Chetyrkin:1981qh}. 
A few progress on a single ladder type topology has been made \cite{Badger:2012dv}, and master integrals for this task have been solved \cite{Henn:2013fah}.

The d-dimensional unitarity method avoids troubles in determining rational terms, but could be much slower than 4-dimensional unitarity method in spinor helicity formalism.
Therefore, it is preferable to use the least possible unitarity cuts for integrand constructions. 
To reduce the number of cuts, we take advantage of the $\mathbb{Z}_4$ symmetry associated with cyclic permutations of external particles. 
Moreover, we choose a set of gauge invariant bases consisting of spins and momenta of external particles and master integrals which makes the $\mathbb{Z}_4$ symmetry manifest, and only compute a minimal set of unitarity cuts which can generate all coefficients of the entire master integrals via the $\mathbb{Z}_4$ symmetry. 

IBP identities are used to reduce the loop integrand to a set of master integrals, and this process is usually a bottleneck of multi-loop computations. 
Considerable and important efforts have been made to enhance the efficiency of IBP reductions \cite{Zeng:2017ipr,Boehm:2018fpv,Bendle:2019csk}.
Recently, with the help of numerical unitarity method, all leading color planar amplitudes and full color helicity-equal amplitudes of two-loop five-gluon scattering process have been achieved \cite{Badger:2015lda,Gehrmann:2015bfy, Badger:2017jhb,Abreu:2017hqn,Abreu:2018jgq,Badger:2018enw, Abreu:2018zmy, Abreu:2019odu,Badger:2019djh}.
\textcolor{black}{Regard to the task at hand, we find that IBP identities with cut conditions imposed \cite{Larsen:2015ped} are quite suitable for the computations}.
IBP reductions with cut conditions are applied directly to cut integrands from unitarity method, where the cut conditions can impressively decrease the number of sectors and master integrals.
This approach not only speed up IBP reductions, but also allows cross checks of the coefficients of master integrals among different cuts.
\textcolor{black}{Moreover, in this approach, the construction of complete loop integrands before IBP reductions can be circumvented, which would be highly non-trivial for a three-loop amplitude.}

Bare helicity amplitudes are obtained by setting the gauge invariant bases to all possible helicity configurations.
Divergences in the ultraviolet-renormalized amplitudes match the known infrared divergence structure \cite{Becher:2009cu, Becher:2009qa} exactly.
Our results provide an essential piece of three-loop amplitudes required for $\textrm{N}^3$LO corrections to di-jet productions in experiments.



\section{Preliminaries and Setups}

The leading color contribution of the four-gluon scattering amplitude can be expressed as a sum of planar amplitudes, dressed with single trace color factors,
\begin{equation}\label{eq:fullamp}
{\bf A}=\sum_{\mathclap{ \sigma\in S_4/\mathbb{Z}_4}} \mathcal {A}(\sigma_1,\sigma_2,\sigma_3,\sigma_4) \textrm{Tr}(T^{a_{\sigma_1}}T^{a_{\sigma_2}}T^{a_{\sigma_3}}T^{a_{\sigma_4}})
\end{equation}
where $S_4/Z_4$ denotes non-cyclic permutations of external particles. 
We only need to compute a particular planar amplitude $\mathcal {A}(1,2,3,4)$, and others can be determined through permutations of kinematic parameters.

Unitarity cut, gauge invariant bases projection and IBP methods are employed to obtain the four-gluon planar amplitude at three loop level. 
The computations are entirely established in the conventional dimension regularization (CDR) scheme, which assumes both internal and external particles are living in $d=4-2\epsilon$ dimensional space.

Cut integrands are constructed using the $d$-dimensional unitarity method. 
The inserted tree amplitudes for cut constructions are written in terms of scalar products of momenta and polarization vectors, and exhibit $d$-dimensional Lorentz invariance. 
The polarization vectors of cutting legs are contracted by the following helicity summing rule
\begin{equation}\label{sumhelicity}
\epsilon_i^{\mu}\circ \epsilon_i^{\nu}\equiv\sum_{\textrm{helicities}} \epsilon_i^{\mu} \epsilon_i^{\nu} = \eta^{\mu\nu} -\frac{p_i^{\mu} q^{\nu} + p_i^{\nu} q^{\mu} }{ q\cdot p_i} \ ,
\end{equation}
with $q$ an arbitrary light-cone reference momentum. 
By this means, cut integrands are manifestly gauge invariant and consist of scalar products of polarization vectors of external gluons, internal and external momenta. 

The external polarizations in the cut integrands can be projected out by the gauge invariant bases method \cite{Boels:2017gyc,Boels:2018nrr} making integrands containing only momenta.  
As shown in \cite{Boels:2018nrr}, the $n$-gluon scattering amplitudes can be expanded by a set of gauge invariant bases, which are constructed by the gauge invariant building blocks $A_i$ and $C_{ij}$
\begin{equation}
\begin{aligned}
A_i =& (\epsilon_i\cdot p_{i+1}) (p_i\cdot p_{i+2}) - (\epsilon_i\cdot p_{i+2}) (p_i\cdot p_{i+1}), \\
C_{ij}= & (\epsilon_i\cdot \epsilon_j) (p_i\cdot p_j ) -  (\epsilon_i\cdot p_j) (\epsilon_j\cdot p_i),\ 
\end{aligned}
\end{equation}
where $i, j \in \{1,2,3,4\}$ and $i+m\sim \textrm{Mod}(i+m,4)$ for $i+m>4$ with $m=1,2$.
For four-gluon scattering amplitudes, we choose the following gauge invariant bases,
\begin{equation}
\begin{aligned} \label{eq:basis}
B_{\alpha} = \Bigl\{&A_1A_2A_3A_4, \ C_{13} C_{24} ,\ C_{13} A_2 A_4,\ \\
&C_{24} A_1 A_3,\ C_{12}C_{34},\ C_{23} C_{14},\  C_{12} A_3 A_4, \\
&\  C_{23} A_4 A_1, \ C_{34} A_1 A_2,\  C_{41} A_2 A_3\ \ \Bigr\}.\\
\end{aligned}
\end{equation}

As a result, the cut integrands can be expressed as
\begin{equation}\label{eq:cutamp}
\begin{aligned}
\mathcal{A}=&\sum_{\alpha} c^{\alpha} B_{\alpha}\equiv \sum_{\alpha}  (\mathcal{A}\circ B^{\alpha})B_{\alpha}\ ,\\
\end{aligned}
\end{equation}
where $B^{\alpha}$ is the dual basis satisfying $B^{\alpha}\circ B_{\beta}=\delta^{\alpha}_{\beta}$ \cite{Boels:2018nrr}.
Polarization vectors are absorbed in $B_{\alpha}$, while the coefficients $c^{\alpha}$ become regular Feynman integrals consisting of the scalar products of internal and external momenta.
\textcolor{black}{One advantage of the gauge invariant bases method is that the integrands are manifestly $d$-dimensional Lorentz invariant and do not contain $\mu^2$-type terms, which are the inner products of the extra $-2\epsilon$ dimensional components of loop momenta. 
These $\mu^2$ terms raise the power of numerators in the integrands, and complicate the Feynman integrals as well as their reductions because of raising powers of numerators of the integrands. 
Moreover, instead of considering different helicity configurations at the beginning of constructing cut amplitudes, gauge invariant bases are running through the entire computations and set to specific helicities in the end.}

The projected cut integrands can be reduced by public IBP programs\cite{vonManteuffel:2012np, Maierhoefer:2017hyi, Maierhofer:2018gpa,Smirnov:2019qkx}, here FIRE6 \cite{Smirnov:2019qkx} is used. 
We apply IBPs with cut constraints directly to the cut integrands, where each unitarity cut channel determines coefficients of a subset of master integrals\cite{Jin:2018fak,Boels:2017gyc}.  

Comparing with the common strategy, in which normal IBPs are used to reduce the complete loop integrands, our new strategy circumvents the reconstruction of the loop integrand from unitarity cut, which can be very difficult in a three-loop computation.
Moreover,  the new strategy allows cross checks of master integral coefficients among different cuts, which is very desirable in a complicated computation.




\section{Unitarity method and Cut-IBP}

\begin{figure}[t]
\includegraphics[scale=0.30]{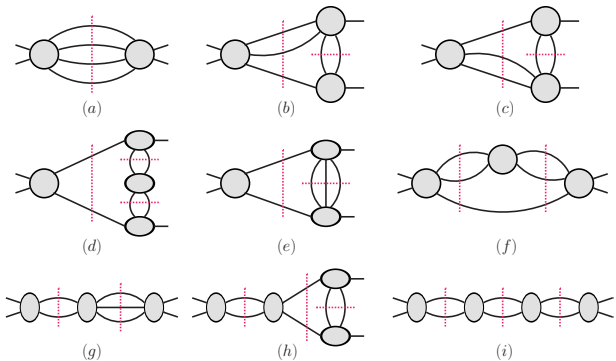}
\caption{A minimal set of unitarity cuts for the 4-gluon 3-loop planar amplitudes.}
\label{fig:allcuts}
\end{figure}

Using the cut-IBP strategy, IBP reduction will be performed for each cut integrand, so we prefer to choose a minmal set of unitarity cuts which can determine the coefficients of all master integrals. 
The number of cuts can be reduced using the $\mathbb{Z}_4$ symmetry of the four-gluon planar amplitude, which is induced by the cyclic permutations of external particles. 
The minimal set of unitarity cuts we choose are enumerated in FIG \ref{fig:allcuts}.


With the help of zone variables \cite{Drummond:2006rz}, a single set of propagators can be used to parameterize all planar diagrams: 
\begin{equation}
\begin{aligned}
&D_i=\{(l_2-l_3)^2,\ (l_1-l_3)^2,\ (l_1-l_2)^2\}\ ,\\
&D_i^a=(l_i+p_1+\cdots +p_{a-1})^2\ ,\\
\end{aligned}\label{props}
\end{equation}
where $i=1,2,3$, $a=1,2,3,4$. 
The coefficients of gauge invariant bases $c^\alpha$ in \eqref{eq:cutamp} after IBP reductions are decomposed into master integrals $\textrm{MI}=\int \prod_{j=1}^{3} dl_j I(D_i,D_i^a)$ and functions of Mandelstam variables $s=(p_1+p_2)^2,\ t=(p_2+p_3)^2$, then the cut amplitudes can be written as
\begin{equation} \label{eq:cutampafterIBP}
\mathcal {A}=\sum_{\alpha}\tilde c^{\alpha}_j(s,t) \textrm{MI}_j B_{\alpha}\ .
\end{equation}



\begin{figure}[t]
\includegraphics[scale=0.25]{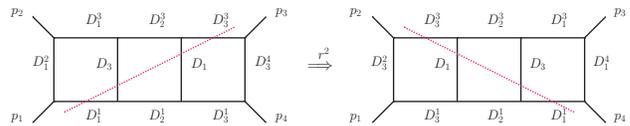}
\caption{Two different parameterizations are used to label one integral. Both parameterizations belong to cut (a) of FIG. \ref{fig:allcuts}. The switch of loop momenta $l_1\leftrightarrow l_3 $ can be induced by a  cyclic permutations $r^2$.}
\label{fig:cyclicinducedredefinition}
\end{figure}

\begin{figure}[htb]
\centering
\includegraphics[scale=0.4]{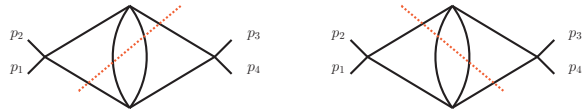}
\caption{A pair of equivalent diagrams with indistinguishable $I(D_i, D_i^a)$ form.}
\label{fig:3loop4ptdiagram513}
\end{figure}

Under a cyclic rotation operation $r$, which is a generator of $\mathbb{Z}_4$ and maps the $i$-th gluon to the $(i+1)$-th,
the Mandelstam variables, propagators and gauge invariant bases in \eqref{eq:cutampafterIBP} transform as
\begin{equation}
\begin{aligned}
r: \ &s\leftrightarrow t,\ \  D_i\rightarrow D_{i+1},\ D_i^a\rightarrow D_{i+1}^a\ ,\\
&B_1\rightarrow B_1,\ B_2\rightarrow B_2,\ 
B_3\leftrightarrow B_4,\ B_5\leftrightarrow B_6,\ \\
&B_7\rightarrow B_8,\ B_8\rightarrow B_9,\ B_9\rightarrow B_{10},
\ B_{10}\rightarrow B_7\ .\\
\end{aligned}\label{z4trans}
\end{equation}
However, the master integrals automatically chosen by FIRE6 do not preserve the $\mathbb{Z}_4$ symmetry. 
We then manually select a new set of integral bases closed under the $\mathbb{Z}_4$ transformations, and transform the old set into the new one.
The cut integrands in terms of new master integral bases are still not manifestly cyclic invariant, due to the fact that the cyclic rotations induce permutations among loop momenta. 
For instance, FIG. \ref{fig:cyclicinducedredefinition} illustrates that twice cyclic rotations induce a switch between loop momenta $l_1\leftrightarrow l_3$, which results in $D_1 \leftrightarrow D_3$ and $D_1^a \leftrightarrow D_3^a$. 
If some master integrands $I(D_i, D_i^a)$ can be mapped to each other by permutations of loop momenta as in FIG. \ref{fig:cyclicinducedredefinition}, they should be identified as the same.
However, if several such equivalent $I(D_i, D_i^a)$ terms are captured by the same cut simultaneously, as illustrated in FIG. \ref{fig:cyclicinducedredefinition} and \ref{fig:3loop4ptdiagram513}, then only one $I(D_i, D_i^a)$ together with its coefficient should be retained, while all the others must be dropped to avoid over counting. 

The consistency between different cuts imposed two conditions to the coefficients of master integrals. First, the coefficients of the equivalent master integrals must be the same. Second, if two integrals are related by the $\mathbb{Z}_4$ symmetry, their coefficients are also related by the $\mathbb{Z}_4$  transformations generated by \eqref{z4trans}. 
Passing these cross checks is a very strong evidence of the correctness of our constructions.

\begin{figure}[htb]
\centering
\includegraphics[scale=0.4]{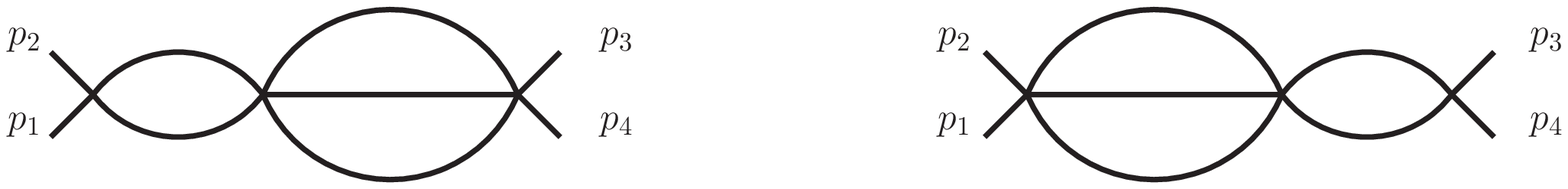}
\caption{A pair diagrams which are topologically different, but correspond to the same integral.}
\label{fig:3loop4ptbubblesunrise}
\end{figure}

Special attention must be paid to a pair of bubble-sunrise type diagrams shown in FIG. \ref{fig:3loop4ptbubblesunrise}.  Although they correspond the same integral, they are distinct diagrams, and their coefficients are related by a $r^2$ transformation. Both the diagrams should be included in the final result.


Another pair of diagrams required more attention, as in FIG. \ref{fig:3loop4ptdiagram513}. 
Similar as FIG. \ref{fig:cyclicinducedredefinition}, these two equivalent diagrams are captured by the same cut
, and only one of them need to be retained.  However, they are indistinguishable in the $I(D_i, D_i^a)$ form, so in the cut integrand the coefficient of this integral is actually twice of its correct value.

If one tries to construct correct loop integrands from unitarity cuts before IBP reductions, all the difficulties and ambiguities above would be exacerbated by the huge number of topologies and higher power numerators therein.



\section{Analytic amplitude}
The analytic forms of master integrals with uniform transcendentality (UT) for planar four-gluon three-loop have been worked out in terms of harmonic polylogarithms (HPL) years ago \cite{Henn:2013fah}.
We transform our symmetric master integral to the UT master integral bases and the bare amplitude can be then expressed in terms of HPLs w.r.t. $\epsilon$ expansion from $\mathcal O(\epsilon^{-6})$ to $\mathcal O(\epsilon^0)$. 

The bare helicity amplitudes are obtained by setting the polarization vectors in the gauge invariant bases to respective helicity configuration. 
As expected, three-loop corrections to the MHV amplitudes start from $\mathcal O(\epsilon^{-6})$, while all-minus and single-plus helicity amplitudes begin with $\mathcal O(\epsilon^{-4})$ .


These bare helicity amplitudes can be expanded as 
\begin{equation}
\mathcal {A}=g_0^2\sum_{L=0}^{\infty}(\frac{\alpha_0}{4\pi})^L C_A^L\mathcal {A}^{(L)}\ ,
\end{equation}
with $\alpha_0=g_0^2/(4 \pi)$ the bare coupling parameters and $C_A$ the quardratic Casmir in the adjoint representation.
Renormalized amplitudes $\mathcal A_R$ can be obtained via replacing bare couplings by renormalized ones derived in $\overline {\textrm{MS}}$ scheme, i.e., 
$\alpha_0=(4\pi)^{-\epsilon} e^{\epsilon \gamma_E}\alpha_s\mu^{2\epsilon} Z_{\alpha}(\alpha,\epsilon)$, and
\begin{equation}
\begin{aligned}
Z_{\alpha}(\alpha,\epsilon)=&1-{\alpha_s\over 4\pi}\frac{ \beta_0}{\epsilon }+{\left(\alpha_s \over 4\pi\right)}^{2}(\frac{\beta_0^{2}}{\epsilon ^{2}}-\frac{\beta_1}{2\epsilon })\\
&+{\left(\alpha_s\over 4 \pi \right)}^{3}(-\frac{\beta_0^{3}}{\epsilon ^{3}}+\frac{7\beta_0\beta_1}{6\epsilon ^{2}}-\frac{\beta_2}{3\epsilon }),\\
\end{aligned}
\end{equation}
where beta function in pure Yang-Mills theory up to three loop level is given by
\bea
\beta_0 = {11\over 3}C_A,\quad \beta_0 = {34\over 3}C_A^2,\quad \beta_0 = {2857\over 54}C_A^3.
\eea

It is also well understood that infrared divergences of renormalized on-shell amplitudes in QCD coincide with the ultraviolet divergences of Wilson coefficients of n-jets operators \cite{Becher:2009cu} in soft-colliear effective theory (SCET) \cite{Bauer:2000yr,Bauer:2001yt,Bauer:2002nz,Beneke:2002ph}. 
These divergences are governed by the infrared renormalization factor $\bf Z$
\bea
{\bf Z}(\{p_i\}, \epsilon,\mu)= {\bf P} \exp{\int_{\mu}^\infty} \frac{d\mu'}{\mu'}{\bf \Gamma} (\{p_i\}, \mu' )\ ,
\eea
where $\bf \Gamma$ has been proposed in \cite{Becher:2009cu}, and for four-gluon planar amplitude it reduces to
\bea
{\bf \Gamma} (s,t;\mu) =\gamma_{\textrm{cusp}} (\alpha_s) \Bigl(\ln {\mu^2\over -s}+\ln {\mu^2\over -t} \Bigr)C_A +4\gamma_g(\alpha_s).\nonumber
\eea
The cusp anomalous dimension $\gamma_{\textrm{cusp}}$ and collinear anomalous dimension $\gamma_g$ up to three loop are listed in \cite{Becher:2009qa} and \cite{Becher:2014oda}. 
Since the $L$-loop correction of ${\bf Z}$ contains at most $1/\epsilon^{2L}$ poles, in order to obtain the three-loop hard function, one- and two-loop amplitudes must be evaluated up to $\mathcal O(\epsilon^4)$ and $\mathcal O(\epsilon^2)$ respectively. 
To this end, We evaluated the one- and two-loop UT master integrals to transcendental degree 6 using differential equations \cite{Henn:2013pwa, Henn:2014qga}, and substituted these solutions into one- and two-loop amplitudes derived under the same scheme \cite{Boels:2017gyc}.

 
Eventually, the hard function is obtained through removing infrared divergence from the renormalized amplitudes,
\begin{equation}\label{hardfunction}
\mathcal {H}=\lim_{\epsilon\rightarrow 0}{\bf Z}^{-1}\mathcal {A}_R\ .
\end{equation}
There is no divergence term with respect to $\epsilon$ remained in $\mathcal {H}$ any more, namely $\mathcal{O}(\epsilon^{-6})$ to $\mathcal{O}(\epsilon^{-1})$ divergences of $\mathcal {A}_R$ must agree with the universal infrared structure exactly. 
This is a strong evidence for the correctness of our result.

In addition, we checked the $s\leftrightarrow t$ symmetry of the amplitudes of helicity configurations $(- - - -)$ and $(+-+-)$.
Besides, amplitude of $(+---)$ helicity also satisfies this symmetry due to its flip symmetry.
The absence of the non-physical pole $\frac{1}{s+t}$ is also checked for analytical forms in both Euclidean and Physical region.

In the ancillary files, bare amplitudes and hard functions in terms of HPL with different helicity configurations are provided, in both Euclidean and Physical regions.



\section{Conclusion and discussion}
In this letter, we have computed the four-gluon three-loop leading color amplitudes in Yang-Mills theory for the first time.
By properly choosing symmetric gauge invariant kinematic bases and master integrals, amplitudes are manifestly invariant under cyclic permutations of external particles.
Coefficients of kinematic bases and master integrals are determined via unitarity cut, projections and cut IBPs.
Hard functions are provided after subtracting UV and IR divergences.
 
Along the same lines, one can further consider all the leading color $\textrm{N}^3$LO four-parton amplitudes in QCD with different initial and final partons.

Our strategy of combining unitarity method and IBP allows crossed checking coefficients of master integrals among different unitarity cuts, and enhances the efficiency of IBP reduction at the same time. 
This strategy can be applied to computations of other multiloop amplitudes.  
For amplitudes with many scales, it should be possible to incorporate the numerical unitarity method \cite{Abreu:2018zmy, Abreu:2019odu, Badger:2019djh}.

\begin{acknowledgments}
\section*{Acknowledgments}
The authors would like to thank  Rutger Boels for initial collaboration on a related project. 
Further more, the authors would like to thank Yanqing Ma and Gang Yang for helpful discussion.
QJ is supported by the National Natural Science Foundation of China (Grants No. 11822508, 11747601), and by the Key Research Program of Frontier Sciences of CAS.
HL is supported by the Recruitment Program of Global Youth Experts of China. 
\end{acknowledgments}

\bibliography{loop_basis.bib}
\end{document}